\documentclass[a4paper, amsfonts, amssymb, amsmath, reprint, showkeys, nofootinbib, twoside,notitlepage,onecolumn]{revtex4-1}

\usepackage{amsmath,amstext}
\usepackage[T1]{fontenc}
\usepackage{amssymb}
\usepackage{graphicx}
\usepackage{ae,aecompl}

\DeclareFontFamily{OT1}{pzc}{}
\DeclareFontShape{OT1}{pzc}{m}{it}{<-> s * [1.10] pzcmi7t}{}
\DeclareMathAlphabet{\mathpzc}{OT1}{pzc}{m}{it}

\usepackage{hyperref}
\usepackage{amsmath}
\usepackage{amssymb}
\usepackage{mathtools}
\usepackage{bm}
\usepackage{cleveref}
\usepackage{tensor}
\usepackage{braket}
\usepackage{enumitem}
\usepackage{mhchem}
\usepackage{amsthm}
\usepackage{nccmath}
\usepackage{mathrsfs}
\usepackage{color}

\newcommand{\E}{{\mathbb{E}}}

\newcommand{\feq}{{f_\textup{eq}}}

\newcommand{\PS}{{\mathbb{P}_S}}
\newcommand{\PF}{{\mathbb{P}_F}}
\newcommand{\s}{{\sigma \!\! \! \sigma}}

\def\be{\begin{equation}}
\def\ee{\end{equation}}
\def\beq{\begin{eqnarray}}
\def\eeq{\end{eqnarray}}

\theoremstyle{definition}

\theoremstyle{theorem}

\theoremstyle{corollary}

\begin{document}
\title{Quasi-hydrodynamics beyond the conventional paradigm}
\author{L.~Gavassino}
\affiliation{Department of Applied Mathematics and Theoretical Physics, University of Cambridge, Wilberforce Road, Cambridge CB3 0WA, United Kingdom}

\begin{abstract}
Quasi-hydrodynamics extends hydrodynamics by promoting slowly relaxing variables to additional degrees of freedom. For such a framework to be rigorously defined and parametrically controlled, the effective theory is usually assumed to retain all the slow modes of the underlying microscopic dynamics, which must be parametrically separated from the rest of the microscopic spectrum and typically arise from weakly broken symmetries. Here, we show that none of these assumptions is essential. Using the effective field theory of quasi-hydrodynamics recently derived from kinetic theory, we study three models that challenge the need for, respectively, parametric scale separation, retaining every slow mode, and a weakly broken symmetry. In all cases, the quasi-hydrodynamic expansion remains fully controlled and can be made arbitrarily accurate. This is possible because the applicability of quasi-hydrodynamics is determined not by the microscopic origin of the slow dynamics, but by the spectral properties of the kinetic operator on a suitably chosen Hilbert space.
\end{abstract} 
\maketitle

\section{Introduction}

The easiest (and most common) way to construct a quasi-hydrodynamic theory is to start from hydrodynamics, and weakly break one of its conservation laws \cite{Grozdanov2019,DavisonGouteraux2016,Delacretaz2017,ArmasTier2023,Brattan2024}. If conservation of a particle number is weakly broken, one obtains the equations of chemical kinetics \cite{BulkGavassino}; if momentum conservation is weakly broken, one obtains Cattaneo-like diffusion \cite{cattaneo1958}; if the elastic stresses are allowed to relax, one recovers Maxwell viscoelasticity \cite{landau7,BAGGIOLI20201} (i.e. Israel--Stewart theory \cite{Israel_Stewart_1979}). In all these examples, the additional quasi-hydrodynamic variables can be viewed as remnants of quantities that would be exactly conserved in an appropriate limit.
Microscopically, this situation arises naturally when a system with an exact symmetry is perturbed by a small symmetry-breaking interaction \cite{Grozdanov2019}. The resulting dynamics has a particularly convenient structure. The smallness of the symmetry-breaking parameter induces a parametric separation between the relaxation rates of the quasi-hydrodynamic variables and those of the remaining microscopic degrees of freedom. Moreover, the slow sector is finite dimensional, and its modes are in one-to-one correspondence with the additional variables retained in the effective theory. Under these conditions, the separation between slow and fast dynamics is unambiguous, and the quasi-hydrodynamic description is parametrically controlled.

Many systems encountered in practice, however, do not fit so neatly into this picture. Kinetic theories are often found to admit remarkably accurate macroscopic descriptions involving only a few moments of the distribution function, even when there is no sharp separation between slow and fast relaxation rates, or when the relevant slow dynamics has no obvious symmetry-related origin \cite{Levermore1984,Ravishankar2010,Denicol2012Boltzmann,Wagner:2023jgq}. By itself, such agreement does not define an Effective Field Theory (EFT). A finite system of moment equations may simply happen to approximate the kinetic evolution well over some range of parameters. In that case, there is no systematic procedure for improving the approximation, and achieving higher accuracy will eventually require introducing additional degrees of freedom.

Here, we are interested in a different possibility: A system may not fall within the standard physical assumptions underlying quasi-hydrodynamics, but nevertheless admit a fully controlled quasi-hydrodynamic effective theory, whose accuracy can be improved systematically without enlarging its set of degrees of freedom. As we shall see, this phenomenon is not exceptional, and can arise in systems realistic enough to be relevant to a wide variety of condensed-matter and astrophysical settings. Our derivation is based on the spectral formulation of quasi-hydrodynamics developed in \cite{GavassinoQuasiEFT:2026ykg}, whose mathematical requirements are considerably weaker than the physical assumptions commonly used to motivate quasi-hydrodynamics.

To illustrate this point, we will study three models of relativistic particles undergoing diffusion through a background medium. Each model violates a different assumption of the conventional quasi-hydrodynamic paradigm:
\begin{itemize}
\item[\textit{Model 1:}] The separation between the slow and fast relaxation timescales is not parametric. Instead, the fast sector relaxes over a timescale only three times shorter than that of the slow sector (see Sec. \ref{sec:model1}).
\item[\textit{Model 2:}] The microscopic theory contains infinitely many modes that relax over the slow timescale, but only one of them needs to be retained in the effective theory (see Sec. \ref{sec:model2}).
\item[\textit{Model 3:}] The parametric separation of the slow mode is unrelated to any weakly broken symmetry, but is tied to the dimensionality and to the detailed structure of the collision integral (see Sec. \ref{sec:model3}).
\end{itemize}
All three systems admit a fully controlled quasi-hydrodynamic description with a fixed number of degrees of freedom.

Throughout the article, we adopt the metric signature $(-,+,+,+)$ and work in natural units, $c=\hbar=k_B=1$.

\section{Quasi-hydrodynamic EFT: a brief recap}\label{quasiEFTgeneral}
\vspace{-0.3cm}

In this section, we briefly review the abstract formalism that allows one to derive a quasi-hydrodynamic EFT from an underlying kinetic theory with a relaxation gap, as first introduced in \cite{GavassinoQuasiEFT:2026ykg}. Here, we will mainly focus on those aspects that are relevant for the present analysis (especially, the convergence domain and truncation error), referring the reader to \cite{GavassinoQuasiEFT:2026ykg} for a more complete discussion.

\vspace{-0.3cm}
\subsection{Formal construction}
\vspace{-0.3cm}

We consider a linear Boltzmann-like equation of the form
\begin{equation}
\partial_t \Psi + \s \Psi + \E^j \partial_j \Psi = 0\, ,
\label{eq:abstract_kinetic}
\end{equation}
where $\Psi:\text{``Minkowski''} \to \mathcal{H}$ is a linearized kinetic degree of freedom taking values in a Hilbert space $\mathcal H$, naturally endowed with the Onsager inner product $(\cdot,\cdot)$, defined so that the local quadratic free-energy density is $\Delta\mathfrak{F}=\frac{1}{2}(\Psi,\Psi)$ (up to, possibly, a constant factor). The operators $\E^j$ and $\s$ are self-adjoint in $\mathcal{H}$, with $\E^j$ the streaming operators and $\s$ a non-negative collision operator. Causality forces $\E^j$ to be bounded. In particular, the maximal speed of propagation $w\equiv \max_{\Omega^j \in S^2}\|\E^j\Omega_j\|$ cannot exceed the speed of light \cite{GavassinoCausality2021,GavassinoLorentzian:2026seq}. Note that, for isotropic systems, $w$ coincides with the spectral norm of each individual $\E^j$.

The quasihydrodynamic EFT assumes the existence of two timescales $\tau_S$ and $\tau_F<\tau_S$ such that $\operatorname{Spectrum}(\s)\subset[0,1/\tau_S]\cup[1/\tau_F,+\infty)$, with $\operatorname{Spectrum}(\s)\cap[0,1/\tau_S]$ consisting of finitely many eigenvalues of finite multiplicity. This spectral separation induces an orthogonal decomposition $\mathcal{H}=\mathcal{H}_S\oplus\mathcal{H}_F$, where $\mathcal{H}_S$ and $\mathcal{H}_F$ are respectively the slow and fast invariant subspaces of $\s$, with associated orthogonal projectors $\PS$ and $\PF=1-\PS$.

Consider a solution of \eqref{eq:abstract_kinetic} of the form $\Psi\propto e^{ik_j x^j-i\omega t}$, corresponding to a quasihydrodynamic mode (i.e.\ a dispersion relation emanating from the slow sector), with $\sqrt{k_jk^j}$ sufficiently small (we will see below how small). Fixing an orthonormal basis $\{e_I\}$ of $\mathcal{H}_S$, we decompose $\Psi=\PS\Psi+\PF\Psi=\sum_H \Psi_H e_H+\Psi_F$. Projecting \eqref{eq:abstract_kinetic} onto $e_I$ and $\mathcal{H}_F$, respectively, gives
\begin{equation}
\label{theboss}
\begin{split}
& -i\omega \Psi_I+\sum_H (e_I,\s e_H)\Psi_H
+ik_j\sum_H(e_I,\E^j e_H)\Psi_H
+ik_j(e_I,\E^j\Psi_F)=0\, , \\
& (\s_F-i\omega+ik_j\E^j_F)\Psi_F
=-ik_j\sum_H \PF\E^j e_H\,\Psi_H\, ,
\end{split}
\end{equation}
where $\s_F=\PF\s\PF$ and $\E_F^j=\PF\E^j\PF$ are compressed operators, understood henceforth as operators on $\mathcal{H}_F$. Since $\operatorname{Spectrum}(\s_F)\subseteq[\tau_F^{-1},+\infty)$, the inverse $\s_F^{-1}$ exists and its norm cannot exceed $\tau_F$. We may therefore write a Neumann expansion $(\s_F{-}i\omega{+}ik_j\E_F^j)^{-1}=[1{-}\s_F^{-1}(i\omega{-}ik_j\E_F^j)]^{-1}\s_F^{-1}=\sum_a[\s_F^{-1}(i\omega{-}ik_j\E_F^j)]^a\s_F^{-1}$, which converges whenever $\|\s_F^{-1}(i\omega-ik_j\E_F^j)\|<1$. Since $\|\s_F^{-1}(i\omega-ik_j\E_F^j)\|\leq\tau_F(|\omega|+w|k|)$ and the quasihydrodynamic frequencies satisfy $|\omega|\leq\tau_S^{-1}+w|k|$, we have $\|\s_F^{-1}(i\omega-ik_j\E_F^j)\|\leq\tau_F/\tau_S+2\tau_Fw|k|<1$ whenever $|k|<(\tau_F^{-1}-\tau_S^{-1})/(2w)$. The Neumann series therefore converges absolutely throughout this domain. We can thus solve the second line of \eqref{theboss} for $\Psi_F$, substitute the result into the first, and transform back to real space. This yields
\begin{equation}
\label{EFTgeneral}
\partial_t \Psi_I
+\sum_H (e_I,\s e_H)\Psi_H
+\sum_H(e_I,\E^j e_H)\partial_j\Psi_H
-\sum_H\sum_{a=0}^\infty
(\PF\E^j e_I,[\s_F^{-1}(-\partial_t-\E^l_F\partial_l)]^a
\s_F^{-1}\PF\E^k e_H)\partial_j\partial_k\Psi_H=0\, ,
\end{equation}
which is a closed system for the slow degrees of freedom $\Psi_I(t,x^j)$ alone. The last term is a gradient expansion that converges within the regime of applicability of the theory, and systematically incorporates the effects of the fast sector without promoting $\Psi_F$ to an independent dynamical degree of freedom.

\vspace{-0.3cm}
\subsection{Convergence domain and truncation error}
\vspace{-0.3cm}

Let us now summarize what assumptions are really needed for \eqref{EFTgeneral} to hold.

First of all, we have seen that the infinite series in \eqref{EFTgeneral} is guaranteed to converge to the \textit{exact result} within the region
\begin{equation}
\label{diaomnd}
|\omega|+w|k|<\tau_F^{-1}\, ,
\end{equation}
and that the quasihydrodynamic dispersion relations are guaranteed to fall in this region for $|k|<(\tau_F^{-1}{-}\tau_S^{-1})/(2w)$. Hence, we do not need $\tau_S$ to be \textit{parametrically} longer than $\tau_F$. It is enough to have a finite gap $\Delta=\tau_F^{-1}-\tau_S^{-1}$ in the spectrum, and a quasihydrodynamic EFT is guaranteed to exist and be controlled within a $k$-range proportional to $\Delta$. Ideally, one would like to have at least $\Delta\sim\tau_S^{-1}$, so that the range of $w|k|$ where the EFT applies is of order $\tau_S^{-1}$.

Additionally, since
$|(\PF\E^j k_j e_I,[\s_F^{-1}(-\partial_t-\E^l_F\partial_l)]^a
\s_F^{-1}\PF\E^k k_k e_H)|
\leq[\tau_F(|\omega|+w|k|)]^a(w|k|)^2\tau_F$,
it is convenient to define the expansion parameter
$q=\tau_F(|\omega|+w|k|)$. Denoting by $R^{(a)}$ the remainder obtained
by truncating the series in \eqref{EFTgeneral} after the term with index
$a$, we have $|R^{(a)}|
\leq
(w|k|)^2\tau_F
\frac{q^{a+1}}{1-q}
\sum_H|\Psi_H|$.
Thus, since $q<1$ throughout the region \eqref{diaomnd}, the truncation
error can be made arbitrarily small at every point strictly inside this
region, even when the gap is not parametrically large. In other words,
parametric scale separation is a convenient way of obtaining a
controlled quasihydrodynamic EFT, but it is not a necessary one.

\vspace{-0.3cm}
\section{Model 1: Cattaneo EFT of small-angle scattering}\label{sec:model1}
\vspace{-0.3cm}

We can now introduce our first ``non-conventional'' quasi-hydrodynamic model. It consists of a kinetic system with a finite relaxation gap, $\operatorname{Spectrum}(\s)\subset[0,1/\tau_S]\cup[1/\tau_F,+\infty)$, but with the ratio $\tau_F/\tau_S=1/3$ fixed and not parametrically small.

\vspace{-0.3cm}
\subsection{Outline of the kinetic model}
\vspace{-0.3cm}

We consider a gas of monoenergetic particles (photons, neutrons, or neutrinos) propagating through a medium at fixed speed $w$. Let \(I(t,x^j,\mathbf n)\) be the angular distribution function, describing the density of particles at the event $(t,x^j)$, and propagating in the direction \(\mathbf n\in S^2\). We regard it as a function \(I:\text{``Spacetime''}\to L^2(S^2)\equiv \mathcal{H}\), endowed at each event with the local inner product \((I,\tilde{I})=\int I^*(\mathbf n)\tilde{I}(\mathbf n)d^2\Omega\). We choose the normalization of \(I\) such that the particle density and flux are given by
\begin{equation}
\rho=\left( \dfrac{1}{\sqrt{4\pi}},I\right) \, , \qquad \qquad J^k=\left( \dfrac{wn^k}{\sqrt{4\pi}},I\right) \, .
\end{equation}
Assuming that the particles do not collide with each other, but scatter off a background medium, the Boltzmann equation in the small-angle elastic scattering limit reads \cite{case1967linear,pomraning1973equations,Levermore1984,mihalas_book}
\begin{equation}\label{boltzmannSmallAngle}
(\partial_t +wn^j \partial_j)I =\dfrac{1}{2\tau} \Delta_{\rm LB} I  \, ,
\end{equation}
where \(\Delta_{\rm LB}\) is the Laplace--Beltrami operator on $S^2$. The collision term on the right-hand side acts on the variable $\mathbf n$, and describes angular diffusion on the momentum sphere (this regime is also known as the Fokker--Planck limit). Comparing \eqref{boltzmannSmallAngle} with \eqref{eq:abstract_kinetic}, we immediately identify $\E^j=wn^j$ and $\s=-(2\tau)^{-1}\Delta_{\rm LB}$, which are manifestly self-adjoint on $\mathcal{H}$. The latter operator is diagonalized by the spherical harmonics $Y^m_\ell$, with eigenvalues $\sigma_\ell=\ell(\ell{+}1)/(2\tau)$. In particular, $\sigma_0{=}0$ corresponds to density conservation, $\sigma_1{=}1/\tau$ to flux relaxation, and $\sigma_2{=}3/\tau$ to stress isotropization.

\vspace{-0.3cm}
\subsection{Cattaneo EFT}
\vspace{-0.3cm}

We retain only the density ($\ell=0$) and the fluxes ($\ell=1$) as dynamical degrees of freedom of the quasihydrodynamic EFT. Hence, we can set $\tau_S=\tau$ and $\tau_F=\tau/3$. The associated gap is $\Delta=2/\tau$, so the EFT is guaranteed to be applicable for $|k| < (w\tau)^{-1}$. This is a conservative bound, and in practice the expansion may converge well beyond it.

The slow Hilbert space $\mathcal{H}_S$ is four-dimensional, and we choose, as orthonormal basis $\{e_I\}$, the vectors $e_0=1/\sqrt{4\pi}$ and $e_j=\sqrt{3}\, n_j/\sqrt{4\pi}$, so that $\Psi_0=(e_0,\Psi)=\rho$ and $\Psi_j=(e_j,\Psi)=\sqrt{3}\, J_j/w$. Then, considering that $e_j=\sqrt{3} n_j e_0$, and therefore $\PF \E^j e_0=0$, the system \eqref{EFTgeneral} explicitly reads
\begin{equation}\label{EFTParticular}
\begin{split}
&\partial_t \rho +\partial_j J^j=0 \, ,\\
& \tau\partial_t J^i+J^i +\dfrac{w^2\tau}{3}\partial^i\rho-w^2\tau\sum_{a=0}^\infty(\PF n^je^i,[\s_F^{-1}(-\partial_t-\E^l_F\partial_l)]^a\s_F^{-1}\PF n^ke^h)\partial_j\partial_k J_h=0 \, .
\end{split}
\end{equation}
We immediately see that, upon dropping the correction series in $a$, the resulting truncated equations correspond to Cattaneo's theory of diffusion \cite{cattaneo1958,rezzolla_book}, with relaxation time $\tau$ and diffusivity $\mathfrak{D}=w^2\tau/3$. The latter coincides with the diffusivity of a random walk in three dimensions, with mean speed $w$ and mean free time $\tau$. The speed of second sound $\sqrt{\mathfrak{D}/\tau}$ in this Cattaneo theory is $w/\sqrt{3}$. Higher orders in the series systematically refine the Cattaneo equation.

\vspace{-0.3cm}
\subsection{First- and second-order corrections}
\vspace{-0.4cm}

The first-order correction in $\tau_F$ comes with the matrix element $(\PF n^je^i,\s_F^{-1}\PF n^ke^h)$. This can be evaluated explicitly by noticing that $\PF n^k e^h$ is a linear combination of $\ell=2$ harmonics, so that $\s_F^{-1}\PF n^ke^h=(\tau/3) \PF n^ke^h$. Hence, we have the following chain of identities:
\vspace{-0.2cm}
\begin{equation}
(\PF n^je^i,\s_F^{-1}\PF n^ke^h)=\frac{\tau}{3}(n^je^i,\PF n^ke^h)=\frac{\tau}{4\pi}(n^jn^i,\PF n^kn^h) =\dfrac{\tau}{15} \left(\delta^{ih}\delta^{jk}+\delta^{ik}\delta^{jh}-\dfrac{2}{3}\delta^{ij}\delta^{hk} \right) \, .
\end{equation}
At second order, the matrix element $(\PF n^je^i,\s_F^{-2}\PF n^ke^h)$ is equal to the first-order result multiplied by an additional factor $\tau/3=\tau_F$, while $(\PF n^je^i,\s_F^{-1}\E_F^l\s_F^{-1}\PF n^ke^h)$ vanishes by parity.
Thus, system \eqref{EFTParticular}, truncated to second order, reduces to
\vspace{-0.2cm}
\begin{equation}\label{EFTParticularsecondorder}
\begin{split}
&\partial_t \rho +\partial_j J^j=0 \, ,\\
& \tau\partial_t J^i+J^i +\mathfrak{D}\partial^i\rho-\left(1-\tau_F\partial_t\right)\dfrac{6}{5}\tau_F\mathfrak{D} \partial_j \left(\partial^{(j} J^{i)}-\dfrac{1}{3}\delta^{ij}\partial_k J^k \right)+\mathcal{O}(\tau_F^3)=0 \, .
\end{split}
\end{equation}
We see that the first-order correction takes the form of an effective shear-viscosity term for the current $J^i$, while the second-order correction introduces a delay $-\tau_F\partial_t$ in the response of this effective viscous stress.

\vspace{-0.5cm}
\subsection{Dispersion relations and accuracy of the EFT}
\vspace{-0.4cm}

To assess the accuracy of the various truncations, we consider solutions $\propto e^{ikx-i\omega t}$, and compare the modes of \eqref{EFTParticularsecondorder} with those of the exact kinetic equation \eqref{boltzmannSmallAngle}. The latter can be computed numerically by observing that, for $I\propto e^{ikx-i\omega t}$, the dispersion relations are determined by $i\omega\in\operatorname{Spectrum}(\s+ik\E^x)$.

\begin{figure}[h!]
    \centering
\includegraphics[width=0.43\linewidth]{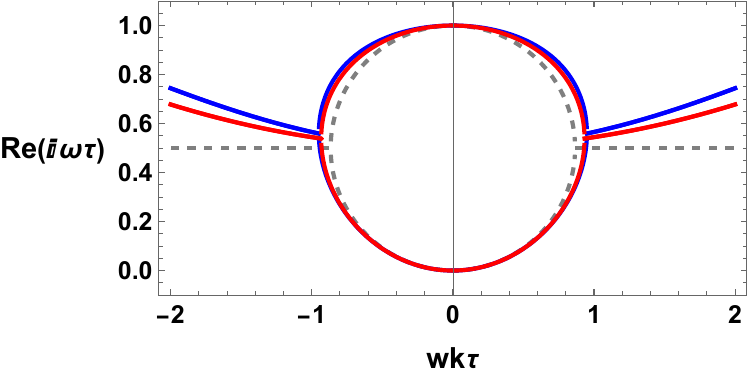}\hspace{0.3cm}\includegraphics[width=0.43\linewidth]{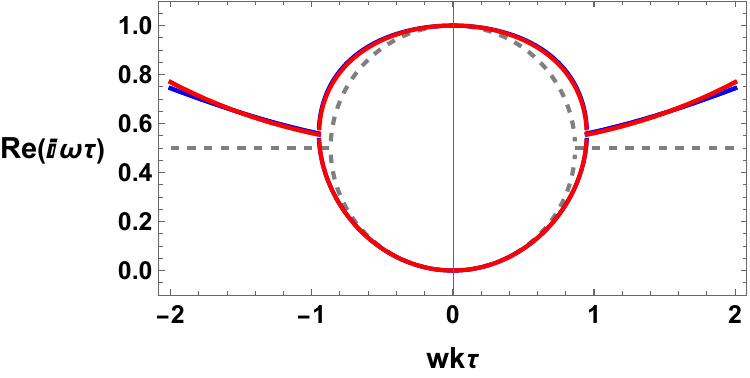}
\includegraphics[width=0.43\linewidth]{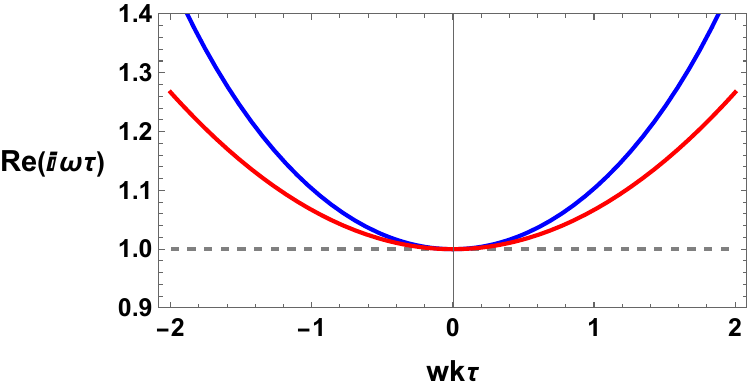}\hspace{0.3cm}\includegraphics[width=0.43\linewidth]{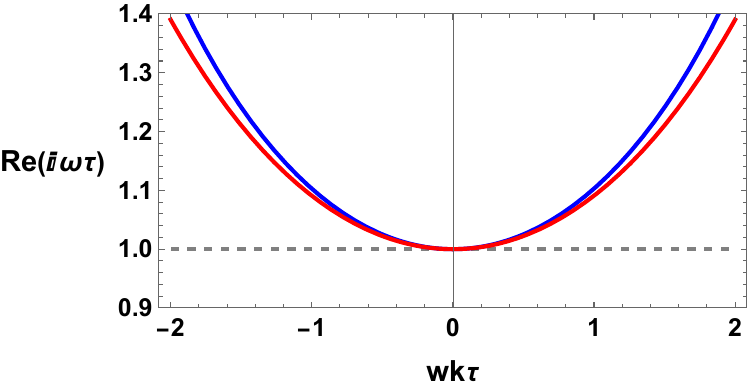}
\caption{Dispersion relations of the longitudinal (upper panels) and transverse (lower panels) quasihydrodynamic modes. We show only $\operatorname{Re}(i\omega\tau)$, since the imaginary parts exhibit a similar pattern and provide no additional information. Blue curves are the numerical solutions of the exact kinetic theory, while gray dashed curves show the zeroth-order (Cattaneo) approximation. Red curves show the first-order approximation in the left panels and the second-order approximation in the right panels. The successive EFT corrections systematically improve the agreement with kinetic theory. In particular, for the longitudinal modes the second-order result is essentially indistinguishable from the exact dispersion relation up to $|wk\tau|\simeq2$, while the agreement is somewhat less accurate for the transverse modes. This deterioration is expected as $|\omega\tau_F|+|wk\tau_F|$ approaches $1$, where the sufficient condition for convergence of the Neumann expansion ceases to hold.}
\label{fig:CattaneoAccuracy}
\end{figure}

The quasihydrodynamic theory supports both longitudinal waves, involving $\rho$ and $J^x$, and transverse waves, involving $J^y$ or $J^z$. Their dispersion relations can be obtained analytically up to second order in the EFT and are compared in Fig.~\ref{fig:CattaneoAccuracy} with the exact kinetic dispersion relations, evaluated numerically.
For longitudinal waves, we find
\vspace{-0.2cm}
\begin{equation}
\begin{split}
&\text{0th order}: \qquad \qquad \qquad\qquad (i\omega\tau)_\pm= \frac{3\pm\sqrt{9-12 (wk\tau)^2}}{6}\, , \\
&\text{1st order}:  \qquad \qquad\qquad\qquad\, (i\omega\tau)_\pm= \frac{45+4 (wk\tau)^2\pm\sqrt{2025-2340 (wk\tau)^2+16 (wk\tau)^4}}{90} \, , \\
&\text{2nd order}: \qquad \qquad \qquad\qquad\!(i\omega\tau)_\pm=\frac{3 \left[45+4 (wk\tau)^2\pm\sqrt{3} \sqrt{675-780 (wk\tau)^2+32 (wk\tau)^4}\right]}{270-8 (wk\tau)^2} \, .
\end{split}
\end{equation}

For transverse waves, instead,
\vspace{-0.2cm}
\begin{equation}\label{shearwaves}
\begin{split}
&\text{0th order}: \qquad \qquad \qquad\qquad i\omega\tau=1 \, , \\
&\text{1st order}:  \qquad \qquad\qquad\qquad i\omega\tau=1+\dfrac{(wk\tau)^2}{15} \, , \\
&\text{2nd order}: \qquad\qquad \!\!\qquad \qquad i\omega\tau= \frac{180}{45-(wk\tau)^2}-3 \, .
\end{split}
\end{equation}
As can be seen, the EFT is already remarkably accurate at second order, especially in the longitudinal channel.

We note that, at the truncation orders considered here, the equations of motion remain of first order in time. There is therefore one mode for each quasihydrodynamic degree of freedom, in agreement with the kinetic theory. At higher orders in the EFT, higher time derivatives eventually appear, generating additional solutions that do not correspond to genuine modes of the underlying dynamics. These spurious modes can be eliminated by order reduction \cite{GavassinoInitialEFT}.

\vspace{-0.4cm}
\subsection{Improved estimate of the UV cutoff scale}
\vspace{-0.3cm}

\begin{figure}[h!]
    \centering
\includegraphics[width=0.43\linewidth]{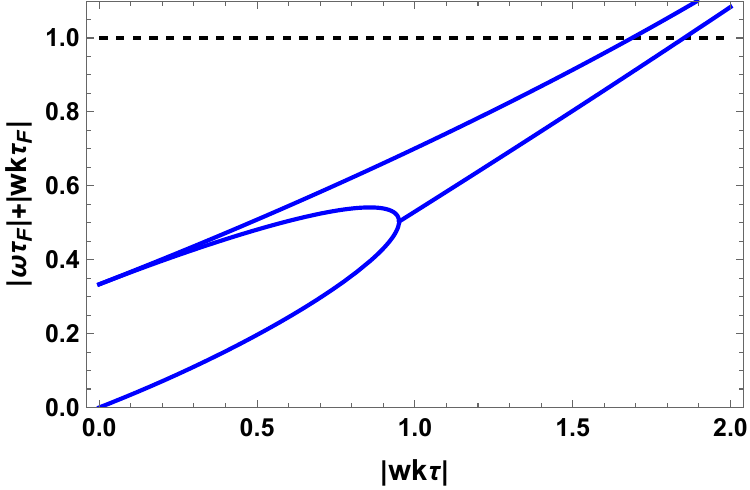}
\caption{Improved estimate of the convergence domain. Blue curves show $|\omega\tau_F|+|wk\tau_F|$ evaluated along all quasihydrodynamic modes of the exact kinetic theory, while the black dashed line marks the threshold $|\omega\tau_F|+|wk\tau_F|=1$. The Neumann expansion is guaranteed to converge as long as all modes remain below this threshold. The first crossing occurs at $|wk\tau|\approx1.7$.}
    \label{fig:TrueCutoff}
\end{figure}

Earlier, we said that the EFT converges to the exact result at least for $|k|<(w\tau)^{-1}$. This estimate assumes the worst-case behavior $|\omega\tau_S|=1+|wk\tau_S|$ for the fastest quasihydrodynamic mode. A sharper estimate can be obtained by evaluating $|\omega\tau_F|+|wk\tau_F|$ along the exact kinetic dispersion relations, since the Neumann series underlying \eqref{EFTParticular} is guaranteed to converge whenever this quantity is smaller than $1$. We find (see Fig.~\ref{fig:TrueCutoff}) that this condition is first saturated at $|wk\tau|\approx1.7$, yielding the improved estimate
\begin{equation}
k_{\text{UV cutoff}}^{\rm Impr.}\approx 1.7(w\tau)^{-1} \, .
\end{equation}
Within this improved convergence domain, the second-order theory already provides an almost exact approximation to the kinetic dispersion relations, at least in the longitudinal sector.

\vspace{-0.3cm}
\subsection{Initial-value formulation and order reduction}
\vspace{-0.4cm}

If we truncate the EFT at a given order, we obtain a system of partial differential equations. For practical applications, this system is only useful if it admits a well-posed and stable initial-value formulation. In \cite{GavassinoQuasiEFT:2026ykg}, it was shown that the zeroth-order theory can always be recast in a causal and stable symmetric-hyperbolic form. Hence, it is well-posed and stable for initial data on arbitrary spacelike hypersurfaces. We also showed that the first-order theory admits a positive definite Lyapunov functional in the rest frame, which implies well-posedness and stability for initial data on equal-time surfaces in that frame. At higher orders, in general, no such result is guaranteed to hold.

Indeed, equation \eqref{shearwaves} reveals that, at second order, the transverse modes become unstable for $(kw\tau)^2>45$, with a growth rate that diverges as $(kw\tau)^2\to45^+$. Thus, if treated as an exact system of equations, the second-order theory is Hadamard ill-posed \cite[\S 3.10]{rauch2012partial}. From an EFT perspective, this is not an inconsistency, since the pathology occurs far beyond the UV cutoff $(wk\tau)^2\approx(1.7)^2$. Still, such a pathology needs to be regularized if the truncated equations are to be used as an initial-value problem. This can be achieved through the usual order-reduction procedure \cite{GavassinoInitialEFT}.
In particular, we can use the lower-order equation $\tau\partial_t J^i+J^i+\mathfrak{D}\partial^i\rho=\mathcal{O}(\tau_F)$ to replace the delay term $\propto\tau_F^2\partial_t$ with terms containing no time derivatives, since the difference between the two can be reabsorbed into $\mathcal{O}(\tau_F^3)$. 

The resulting order-reduced theory is
\begin{equation}
\begin{split}
&\partial_t \rho +\partial_j J^j=0 \, ,\\
& \tau\partial_t J^i+J^i +\mathfrak{D}\partial^i\rho-\dfrac{\tau\mathfrak{D}}{15} \left[\partial_j\partial^{j} \left(4J^{i}+\mathfrak{D}\partial^i \rho\right)+\dfrac{1}{3}\partial^i\partial_j \left(4J^j+\mathfrak{D}\partial^j \rho\right) \right]+\mathcal{O}(\tau_F^3)=0 \, .
\end{split}
\end{equation}
One can verify explicitly that the order-reduced dispersion relations are as accurate as those of the original second-order EFT up to $|wk\tau|\simeq2$, while remaining rest-frame stable at arbitrarily large wavenumbers. Thus, order reduction preserves the accuracy of the EFT in its regime of applicability while restoring a stable and well-posed initial-value formulation in the rest frame. 

\vspace{-0.4cm}
\section{Model 2: Two-temperature EFT of scattering-dominated radiative transfer}\label{sec:model2}
\vspace{-0.3cm}

We now introduce our second model. It describes a system whose slow sector is infinite dimensional, but only one slow mode couples to the conservation law. The remaining slow modes can then be excluded by appropriately restricting the Hilbert space $\mathcal{H}$.

\vspace{-0.4cm}
\subsection{Outline of the kinetic model}
\vspace{-0.3cm}

We consider a radiative-transfer model consisting of a photon gas coupled to a rigid gray medium in local thermodynamic equilibrium. The medium has heat capacity per unit volume $c_M$ and temperature perturbation $\delta T_M$, while the photon gas is described by the perturbation $\delta f(t,x^j,p^j)$ of its distribution function, with $p^\mu$ the photon four-momentum. The linearized equations of radiative transfer are \cite{pomraning1973equations,mihalas_book,Levermore1984,GavassinoRadiativeBounds:2025bxx}
\begin{equation}\label{radiativetransferequaitoningeneral}
\begin{split}
& c_M \partial_t \delta T_M
+ \int \dfrac{2d^3 p}{(2\pi)^3} p^0
(\partial_t {+}n^j \partial_j )\delta f =0 \, , \\ 
&(\partial_t {+}n^j \partial_j )\delta f
= \dfrac{\feq(1{+}\feq)\beta^2 p^0 \delta T_M-\delta f}{\tau_\text{abs}}
+\dfrac{\langle \delta f \rangle-\delta f}{\tau_\text{sc}} \, ,
\end{split}
\end{equation}
where $\mathbf n=\mathbf p/p^0\in S^2$, $\beta=1/T$, and $\feq=(e^{\beta p^0}-1)^{-1}$ is the equilibrium black-body distribution. The first equation expresses conservation of the total energy of matter and radiation. In the second, $\tau_\text{abs}$ and $\tau_\text{sc}$ are the absorption-emission and scattering timescales, which are momentum independent by the gray-medium assumption. The absorption term, fixed by the Kirchhoff-Planck relation \cite{mihalas_book}, drives the photon gas toward a black-body distribution at the local matter temperature $T+\delta T_M$. Scattering is instead elastic and isotropic, with $\langle\cdot\rangle$ denoting the angular average, and therefore isotropizes the photon distribution without exchanging energy with the medium.

\vspace{-0.4cm}
\subsection{Separation of timescales}
\vspace{-0.3cm}

We assume that scattering is much faster than absorption, $\tau_\text{sc}\ll\tau_\text{abs}$, so that $\tau_F\sim\tau_\text{sc}$. The anisotropic part $\delta f-\langle\delta f\rangle$ is then fast, while the slow sector remains infinite dimensional because every isotropic distribution relaxes only on the absorption timescale $\tau_S \sim \tau_\text{abs}$. Crucially, however, only one of these slow degrees of freedom carries energy and can couple to the matter temperature.
To see this, consider the general decomposition
\begin{equation}
\delta f=\feq(1{+}\feq)\beta^2p^0 \sum_{n=0}^{\infty} \chi_n(p^0)I_n(t,x^j,\mathbf{n}) \, .
\end{equation}
We fix $\chi_0=1$, and choose $\chi_n$ to be Onsager-orthogonal. In particular, we require the quadratic free-energy-density perturbation to decouple as a sum of contributions of the form $\beta c_R\langle I_n^2\rangle$, with $c_R$ the heat capacity per unit volume of the photon gas. This gives
\begin{equation}\label{reducethefreenergyofradiation}
\begin{split}
2\Delta\mathfrak{F}={}& \beta c_M (\delta T_M)^2 +\int \dfrac{2d^3 p}{(2\pi)^3} \dfrac{(\delta f)^2}{\feq (1{+}\feq)\beta}\\
={}& \beta c_M (\delta T_M)^2 +\beta\sum_{n,m}\langle I_n I_m\rangle\int \dfrac{2d^3 p}{(2\pi)^3} \feq(1{+}\feq)(\beta p^0)^2 \chi_n \chi_m \\
\stackrel{!}{=}{}& \beta c_M (\delta T_M)^2 +\beta c_R\sum_{n}\langle I_n^2 \rangle \qquad  \qquad \Longrightarrow \qquad \qquad \int \dfrac{2d^3 p}{(2\pi)^3} \feq(1{+}\feq)(\beta p^0)^2 \chi_n \chi_m=c_R \delta_{nm} \, .
\end{split}
\end{equation}
With this choice of basis, $\langle I_0\rangle$ coincides with the Landau-matched temperature perturbation of the photon gas:
\begin{equation}
\delta T_R \equiv\dfrac{\delta T^{00}_R}{c_R}=\dfrac{1}{c_R}\int \dfrac{2d^3p}{(2\pi)^3} p^0 \delta f=\dfrac{1}{c_R}\sum_{n=0}^{\infty}\int \dfrac{2d^3p}{(2\pi)^3} \feq(1{+}\feq)(\beta p^0)^2 \chi_n\chi_0 \langle I_n \rangle =\langle I_0 \rangle \, .
\end{equation}
\newpage
As a result, the system \eqref{radiativetransferequaitoningeneral} reduces to
\vspace{-0.2cm}
\begin{equation}\label{radiativetransferequaitonreduce}
\begin{split}
& c_M \partial_t \delta T_M
 = \dfrac{c_R(\delta T_R- \delta T_M)}{\tau_\text{abs}}
 \, , \\ 
&(\partial_t {+}n^j \partial_j )I_0
= \dfrac{\delta T_M-I_0}{\tau_\text{abs}}
+\dfrac{\delta T_R- I_0}{\tau_\text{sc}} \, ,\\
&(\partial_t {+}n^j \partial_j ) I_n
= -\dfrac{I_n}{\tau_\text{abs}}
+\dfrac{\langle  I_n \rangle- I_n}{\tau_\text{sc}} \qquad \qquad (n\neq 0)\, .\\
\end{split}
\end{equation}
Hence, all degrees of freedom $\{I_n\}_{n\neq 0 }$ with $n\neq 0$ completely decouple from the evolution of $\delta T_M$ and $I_0$. Therefore, despite $\{\langle I_n \rangle\}_{n\neq 0}$ being formally slow degrees of freedom, we can ignore them, and construct a quasi-hydrodynamic EFT involving only $\delta T_M$ and $\delta T_R =\langle I_0 \rangle$. This is accomplished following the same procedure as in Sec. \ref{quasiEFTgeneral}, but working within an appropriate Hilbert space $\mathcal{H}$ that contains only $\delta T_M$ and $I_0$, as we show below.

\vspace{-0.5cm}
\subsection{Two-temperature EFT}
\vspace{-0.4cm}

\begin{figure}[h!]
    \centering
\includegraphics[width=0.42\linewidth]{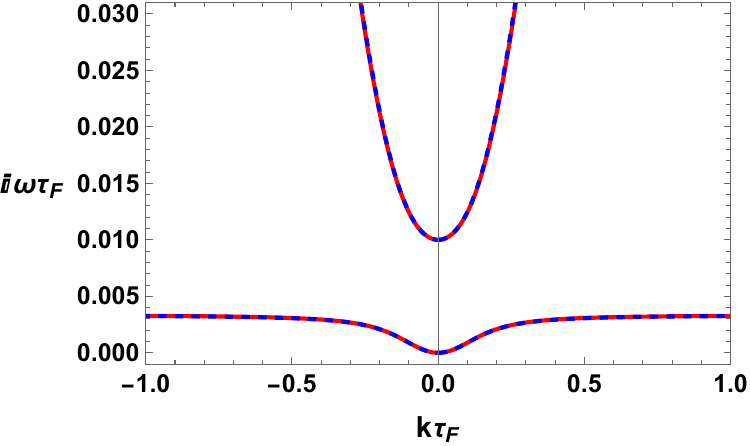}\hspace{0.3cm}\includegraphics[width=0.42\linewidth]{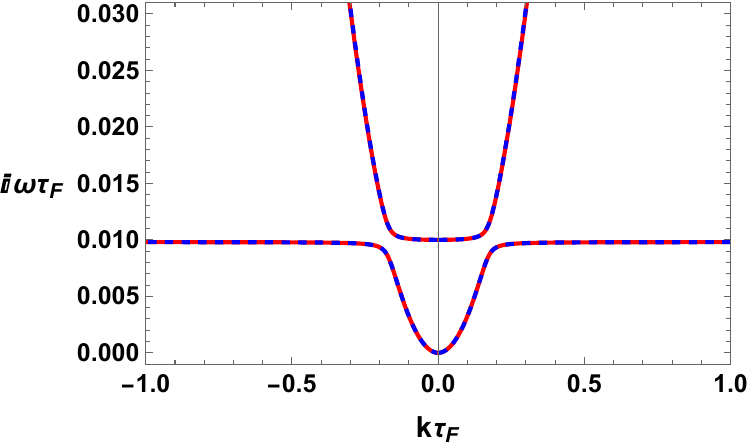}
\caption{Dispersion relations of the two-temperature model for $\tau_S=100\,\tau_F$, in the matter-dominated regime $c_R/c_M=0.5$ (left) and in the radiation-dominated regime $c_R/c_M=50$ (right). The blue dashed curves are the exact quasihydrodynamic modes of the kinetic theory, obtained analytically from \eqref{radiativetransferequaitonreduce}, while the red solid curves are the predictions of the first-order EFT \eqref{lunchEFT}. Frequencies and wave numbers are expressed in units of $\tau_F^{-1}$. Despite being derived as a low-frequency and long-wavelength expansion, the first-order EFT remains remarkably accurate throughout the range shown, including at $k\tau_F\sim 1$.}
\label{fig:TwoTemp}
\end{figure}

The first two lines of \eqref{radiativetransferequaitonreduce} can be recast in the canonical symmetric form \eqref{eq:abstract_kinetic} once we Onsager-normalize its degrees of freedom. In particular, from \eqref{reducethefreenergyofradiation}, we see that, if we define 
\begin{equation}
\Psi=
\begin{bmatrix}
\Psi_T \\
\psi
\end{bmatrix}
=
\begin{bmatrix}
\sqrt{\beta c_M} \,\delta T_M \\
\sqrt{\beta c_R} \, I_0
\end{bmatrix} \, ,
\end{equation}
then we can set $\mathcal{H}=\mathbb{C}\oplus L^2(S^2)$, equipped with the Onsager inner product $(\Phi,\Psi)=\Phi_T^*\Psi_T+\langle\phi^*\psi\rangle$. Thus, we recover equation \eqref{eq:abstract_kinetic}, with
\vspace{-0.4cm}
\begin{equation}
\E^j
\begin{bmatrix}
\Psi_T\\
\psi
\end{bmatrix}
=
\begin{bmatrix}
0\\
n^j\psi
\end{bmatrix} \, , \qquad\qquad
\s
\begin{bmatrix}
\Psi_T\\
\psi
\end{bmatrix}
=
\begin{bmatrix}
\dfrac{1}{\tau_\text{abs}}
\left(
\dfrac{c_R}{c_M}\Psi_T
-\sqrt{\dfrac{c_R}{c_M}}\langle\psi\rangle
\right)
\\[2mm]
\dfrac{1}{\tau_\text{abs}}
\left(
\psi-\sqrt{\dfrac{c_R}{c_M}}\Psi_T
\right)
+\dfrac{\psi-\langle\psi\rangle}{\tau_\text{sc}}
\end{bmatrix} \, .
\end{equation}
Both operators are self-adjoint in $\mathcal{H}$. We immediately find that $\s$ has eigenvector $[\sqrt{c_M},\sqrt{c_R}]^T$ with eigenvalue $0$, and another eigenvector $[-\sqrt{c_R},\sqrt{c_M}]^T$ with eigenvalue $\tau_\text{abs}^{-1}(1+c_R/c_M)$. These two vectors define the slow sector, so we can set $\tau_S=\tau_\text{abs}/(1+c_R/c_M)$. The rest of the spectrum sits at $\tau_F^{-1}=\tau_\text{abs}^{-1}+\tau_\text{sc}^{-1}\approx\tau_\text{sc}^{-1}$. The resulting first-order quasihydrodynamic EFT, expressed in non-normalized variables, then reads
\vspace{-0.2cm}
\begin{equation}\label{lunchEFT}
\begin{split}
\partial_t \delta T_M
={}&\dfrac{c_R(\delta T_R-\delta T_M)}{c_M\tau_\text{abs}} \, , \\ 
\partial_t \delta T_R
={}&\dfrac{\delta T_M-\delta T_R}{\tau_\text{abs}}
+\frac{\tau_F}{3}\partial^j\partial_j\delta T_R
+\mathcal{O}(\tau_F^2) \, .
\end{split}
\end{equation}
Remarkably, when $\tau_F\ll\tau_S$, this first-order theory remains very accurate even at arbitrarily large $k$, provided that $\omega\tau_F\ll1$ (see Fig.~\ref{fig:TwoTemp}). In particular, the hydrodynamic mode approaches a constant frequency at large $k$, with the exact theory and the EFT reproducing the same asymptote. 

Physically, this EFT describes a two-temperature system \cite{GavassinoRadiazione}, with $\delta T_R$ the Landau-matched temperature perturbation of the radiation gas. Importantly, this description remains valid even when the full $p^0$-dependence of $\delta f$ is highly nonthermal. Indeed, the infinitely many additional slow excitations $\{I_n\}_{n\neq 0}$ carry no energy and decouple from the dynamics of $\delta T_M$ and $\delta T_R$. This mechanism is not specific to radiative transfer. More generally, a mode may relax arbitrarily slowly and yet be irrelevant to a given effective description if it belongs to a sector that does not couple dynamically to the observables of interest. Such decoupled sectors can arise, for example, from selection rules or invariant subspaces of the microscopic dynamics, while the choice of observables determines which of these sectors are relevant. Once such irrelevant sectors have been identified, they can be rigorously excluded by appropriately restricting the Hilbert space $\mathcal H$ within which the quasihydrodynamic EFT is constructed.

\section{Model 3: Cattaneo EFT of bottlenecked momentum diffusion in one dimension}\label{sec:model3}

We now introduce our third model. It describes a $(1+1)$-dimensional Cattaneo-like system in which the particle current is quasi-conserved, but this quasi-conservation law arises ``by accident'', as a consequence of the low dimensionality and the structure of the collision integral. In particular, there is no obvious approximate symmetry underlying the quasi-conservation of the current, such as weakly broken translation invariance or the phase dynamics of a superfluid.

\subsection{Outline of the kinetic model}

We consider a non-degenerate gas of massless probe particles constrained to move on a line. The perturbation to their distribution function is $\delta f(t,x,p)$, where $p$ is the particle momentum. The energy of each particle is $\varepsilon=|p|$, while its velocity is $v=\text{sign}(p)$. These particles scatter off a heat bath held at fixed inverse temperature $\beta$. In the limit of soft but frequent scatterings, the perturbation obeys the Vlasov--Fokker--Planck equation \cite{Debbasch1997,Dunkel:2008ngc,GavassinoFokkerPLanck:2026zsz,GavassinoDiffusionCompatible:2026tvy}
\begin{equation}\label{vlasovfokkerplanck}
(\partial_t +v\partial_x) \delta f
=\partial_p [\mathfrak{B}(p)(\partial_p \delta f +\beta v \delta f)] \, .
\end{equation}
Often, one assumes that the momentum diffusivity $\mathfrak{B}(p)$ is constant. Here, instead, we assume that scattering is very efficient at high energies (compared with the temperature), and much less efficient at low energies. To capture this behavior, we set $\mathfrak{B}(\infty)=4/(\beta^2\tau_F)$, where we shall see that $\tau_F$ is precisely the fast timescale, and $\mathfrak{B}(0)=4/(\beta^2\tau_0)$, with $\tau_0\gg\tau_F$. As we shall see, the slow timescale $\tau_S$ is related to $\tau_0$. A simple interpolating model is
\begin{equation}\label{ourchoiceofB}
\mathfrak{B}(p)= \dfrac{4}{\beta^2\tau_F} \left[1-\left(1{-}\sqrt{\dfrac{\tau_F}{\tau_0}}\right)e^{-\beta \varepsilon}\right]^2 \, .
\end{equation}
The exact analytical form of $\mathfrak{B}(p)$ is not important for the existence of the Cattaneo EFT, but this specific choice allows us to determine the full spectrum analytically.

\subsection{Onsager-normalized representation}

If we Onsager-normalize the degrees of freedom, the collision operator in equation \eqref{vlasovfokkerplanck} becomes Schr\"odinger-like. In particular, setting $\delta f=\sqrt{2\pi}\,\mathfrak{B}^{-1/4}e^{-\beta\varepsilon/2}\Psi$ and $dq=\mathfrak{B}^{-1/2}dp$, the quadratic free-energy-density perturbation becomes
\begin{equation}
2\Delta\mathfrak{F}=\int \dfrac{dp}{2\pi} e^{\beta\varepsilon}(\delta f)^2 =\int \Psi^2 dq \, , 
\end{equation}
so that we can set $\mathcal{H}=L^2(\mathbb{R})$, equipped with the usual inner product $(\Phi,\Psi)=\int \Phi^* \Psi \, dq$. Moreover, solving the differential equation $dq=\mathfrak{B}^{-1/2}dp$, we obtain 
\begin{equation}
q=\dfrac{\sqrt{\tau_F}\, \text{sign}(p)}{2} \ln \left[1+\sqrt{\dfrac{\tau_0}{\tau_F}}(e^{\beta\varepsilon}{-}1)\right] \qquad  \Longrightarrow  \qquad \mathfrak{B}(q)=\dfrac{4}{\beta^2 \tau_F} \left[1+\left(\sqrt{\dfrac{\tau_0}{\tau_F}}-1\right)e^{-2|q|/\sqrt{\tau_F}}\right]^{-2} .
\end{equation}
After a tedious calculation, equation \eqref{vlasovfokkerplanck} acquires the general form \eqref{eq:abstract_kinetic}, where the relevant operators are
\begin{equation}
\E=\text{sign}(q) , \qquad \s=-\partial^2_q-\dfrac{2\tanh(g)}{\sqrt{\tau_F}} \delta(q)+\dfrac{1}{\tau_F} \left[1-2\,\text{sech}^2 \left(\dfrac{|q|}{\sqrt{\tau_F}}+g \right) \right] ,\qquad \text{with }e^{-2g}=\sqrt{\dfrac{\tau_0}{\tau_F}} -1  .
\end{equation}
Both operators are self-adjoint in the appropriate domains, and $\E$ has norm $1$. Different choices of $\mathfrak{B}$ would still result in $\s$ being an effective Schr\"odinger Hamiltonian, but the potential would be different.

\subsection{Exact spectrum of the model}
\vspace{-0.3cm}

The excitations of the model are defined by the condition $i\omega\in \text{Spectrum}(\s +ik \E)$, i.e. $i\omega$ is an eigenenergy of the Hamiltonian $\s$ perturbed by the potential $ik \E$ (which is anti-Hermitian on Fourier modes). This can be determined analytically, as we show below.

\begin{figure}[h!]
    \centering
\includegraphics[width=0.39\linewidth]{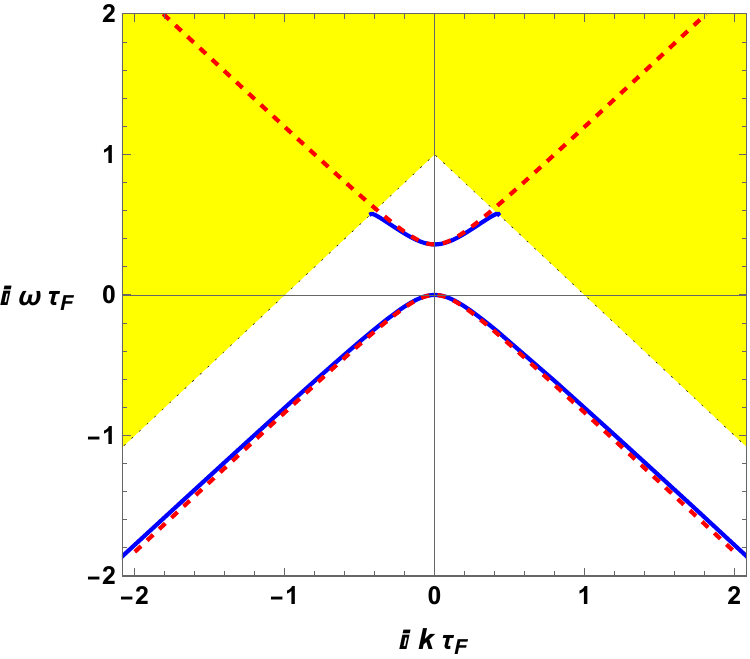}\hspace{0.3cm}\includegraphics[width=0.39\linewidth]{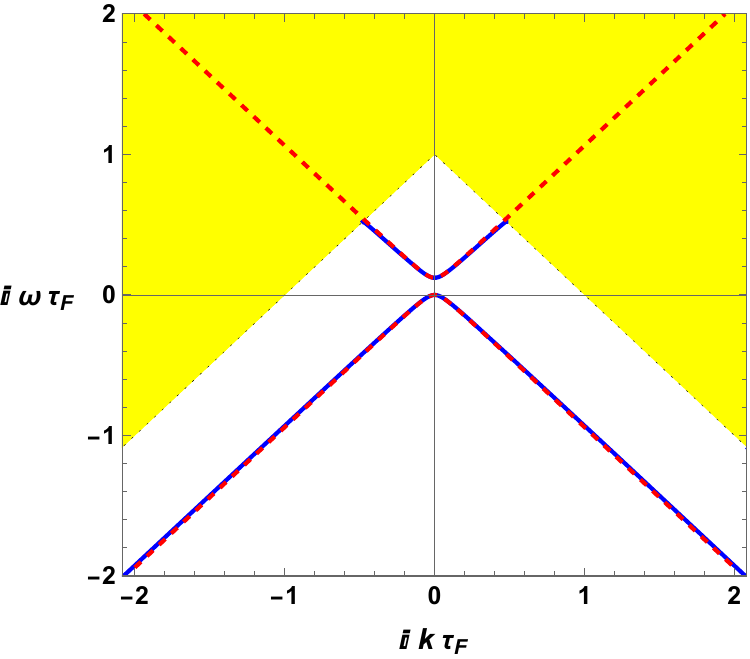}
\caption{Excitation spectrum of the model \eqref{vlasovfokkerplanck}-\eqref{ourchoiceofB} at imaginary wavenumber, for $\tau_0\,{=}\,10^2\tau_F$ (left panel) and $10^3\tau_F$ (right panel). The yellow region is a continuous ballistic branch, which is treated here as the fast sector. The blue dispersion relations are the exact quasi-hydrodynamic modes. We choose to work at imaginary $k$ because this allows us to solve \eqref{1ddispersiamo} analytically as a parametric curve with real parameter. In the limit $\tau_F \ll \tau_S$, the model is increasingly well approximated near the origin by the Cattaneo theory (red dashed).}
\label{fig:Fokker}
\end{figure}

The essential spectrum of Schr\"odinger operators can be determined by analyzing the large-$q$ limit (Weyl criterion). Here,  $\s+ik\E$ becomes the Hamiltonian of a free particle, shifted by the potential $1/\tau_F\pm ik$. Hence, the excitation spectrum contains two continuous branches, $i\omega=\pm ik+[\tau_F^{-1},\infty)$,
which correspond to high-energy degrees of freedom that propagate ballistically through the medium. To determine the discrete eigenvalues, instead, we need to solve the Schr\"odinger equation explicitly:
\vspace{-0.3cm}
\begin{equation}
\begin{split}
&\tau_F\partial^2_q\Psi+2 \,\text{sech}^2 \left(g-\tfrac{q}{\sqrt{\tau_F}} \right)\Psi= (1-i\omega\tau_F-ik\tau_F)\Psi \qquad \qquad (\text{for }q<0)\, ,\\
&\tau_F\partial^2_q\Psi+2 \text{sech}^2 \left(g+\tfrac{q}{\sqrt{\tau_F}} \right) \Psi= (1-i\omega\tau_F+ik\tau_F)\Psi \qquad \qquad (\text{for }q>0)\, ,\\
& \sqrt{\tau_F}\, \Psi'(0^+)-\sqrt{\tau_F}\, \Psi'(0^-)= - 2\tanh(g)\Psi(0)  \, . \\
\end{split}
\end{equation}
Solving on the two sides (retaining the solutions that decay at infinity), and imposing continuity at $q=0$, we obtain
\vspace{-0.2cm}

\begin{equation}
\Psi(q)=\text{const} \times \begin{cases}
(r_++\tanh\left(g\right)) \left[r_-+\tanh\left(g{-}\frac{q}{\sqrt{\tau_F}}\right)\right] e^{+\frac{r_- q}{\sqrt{\tau_F}}}\, , & q<0\, , \\[3mm]
(r_- +\tanh\left(g\right))\left[r_+ +\tanh\left(g{+}\frac{q}{\sqrt{\tau_F}} \right)\right] e^{-\frac{r_+ q}{\sqrt{\tau_F}}}  \, , & q>0\, , \\
\end{cases}
\end{equation}
with $r_\pm=\sqrt{1-i\omega\tau_F\pm ik\tau_F}$, where we take the root with positive real part. Imposing the matching condition on the derivative at the origin, we obtain
\vspace{-0.2cm}
\begin{equation}\label{1ddispersiamo}
(r_+^2-1)\left[r_-+\tanh(g)\right]+(r_-^2-1)\left[r_++\tanh(g)\right]=0 \, .
\end{equation}
This is an implicit function in $\omega$ and $k$, which provides the \textit{exact} dispersion relations of all the discrete modes of the model (with exceptional points understood by continuity). In the $k\to 0$ limit, where $r_+$ and $r_-$ coincide, we obtain a hydrodynamic mode, $i\omega=0$, and a single slow non-hydrodynamic frequency:
\vspace{-0.2cm}
\begin{equation}
i\omega=\dfrac{4}{\sqrt{\tau_F \tau_0}}\left[1-\sqrt{\dfrac{\tau_F}{\tau_0}}\right] \, .
\end{equation}
Hence, in the limit $\tau_F/\tau_0\to 0$, we identify $\tau_S\approx \sqrt{\tau_F \tau_0}/4$, and \eqref{1ddispersiamo} becomes the dispersion relation of Cattaneo's theory, $\tau_S\omega^2+i\omega-\tau_Sk^2+\mathcal{O}(\tau_F)= 0$
thereby establishing that the present model is a Cattaneo-type quasi-hydrodynamic system (see figure \ref{fig:Fokker}).

\subsection{Origin of the quasi-hydrodynamic behavior}
\vspace{-0.3cm}

We can understand the emergence of a Cattaneo-like effective behavior through the following argument. 

Let $\rho_\pm=\int_{\mathbb{R}_{\pm}} \frac{dp}{2\pi} \delta f$ denote the densities of left- and right-moving particles. Integrating equation \eqref{vlasovfokkerplanck}, we obtain
\begin{equation}
\begin{split}
& (\partial_t-\partial_x)\rho_- =+ \dfrac{\mathfrak{B}(0)}{2\pi} \feq \partial_p \left( \delta f/\feq \right)_{p=0} \, ,\\
& (\partial_t+\partial_x)\rho_+ = - \dfrac{\mathfrak{B}(0)}{2\pi} \feq \partial_p \left( \delta f/\feq \right)_{p=0}\, .\\
\end{split}
\end{equation}
In the limit where $\mathfrak{B}(0)\to 0$, the exchange between these two populations is suppressed, causing them to decouple and become purely propagating degrees of freedom. As a result, not only is the total density $\rho_++\rho_-$ conserved, but also the total current $\rho_+-\rho_-$. At finite but small $\mathfrak{B}(0)$, the current is no longer conserved, but it relaxes parametrically slowly due to bottlenecked momentum diffusion near $p=0$.

\begin{figure}[h!]
    \centering
\includegraphics[width=0.41\linewidth]{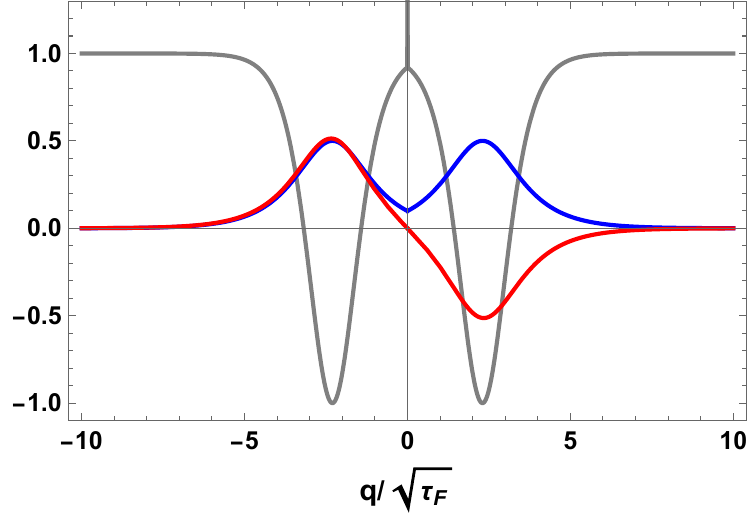}\hspace{0.6cm}\includegraphics[width=0.41\linewidth]{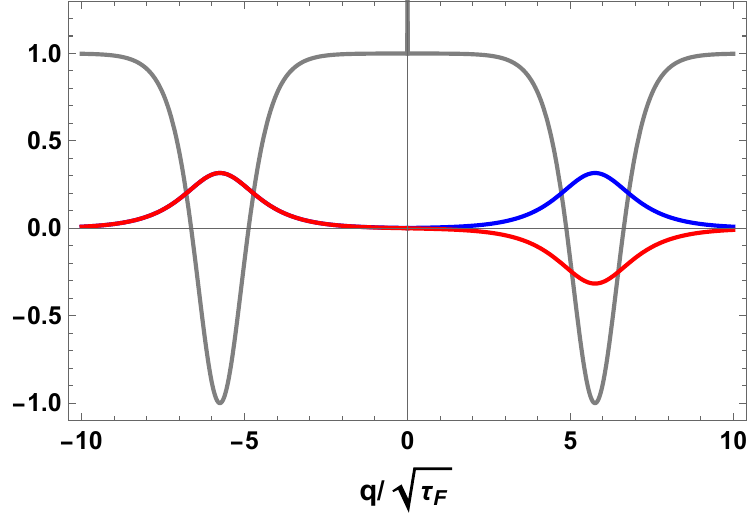}
\caption{Effective potential (gray) of the model \eqref{vlasovfokkerplanck}-\eqref{ourchoiceofB} for $\tau_0=10^4 \tau_F$ (left panel) and $10^{10}\tau_F$ (right panel). The blue curve is the effective wavefunction of the hydrodynamic mode, which corresponds to the ground state of the quantum system, while the red curve is the effective wavefunction of the quasi-conserved current mode, which corresponds to the only excited bound state of the quantum system. At large $\tau_0/\tau_F$, the two potential wells move progressively further away, and the two wavefunctions reduce to respectively the even and odd linear combinations of the P\"oschl--Teller bound states localized within each well. As a result, their energy separation tends to zero, while the rest of the spectrum lies in the continuum, making the emergence of a Cattaneo EFT possible.}
\label{fig:EffPot}
\end{figure}

The above argument explains why the current is quasi-conserved when $\tau_0\gg \tau_F$. However, it does not explain why it is the \textit{only}
quasi-conserved degree of freedom, since in principle there could be other low-energy observables that become parametrically slow as $\mathfrak{B}(0)\to0$. This is where it becomes important to examine the detailed structure of the relaxation operator $\s$. For $\tau_0 \gg \tau_F$, the latter is the Hamiltonian of a quantum particle in a symmetric double-well potential, supplemented by a repulsive Dirac-delta barrier at the origin (see Fig. \ref{fig:EffPot}). In the limit $\tau_0/\tau_F \to \infty$, the two wells move infinitely far away from each other. Each isolated well has the P\"oschl--Teller form and supports a single bound state, while all remaining states belong to the continuum. Hence, for sufficiently large $\tau_0/\tau_F$, the full potential possesses only two bound states below the continuum, approximately corresponding to the even and odd linear combinations of the P\"oschl--Teller bound states localized in the two wells, as shown in Fig. \ref{fig:EffPot}. The even state is the equilibrium zero mode, which, being even in momentum, carries a total particle density but no current. The odd state is the quasi-conserved current mode, which, being odd, carries no total particle density but a finite current. The energy splitting $1/\tau_S$ between the two states is the inverse of the tunneling time between the two wells, and shrinks to zero as $\tau_0/\tau_F$ tends to infinity. All other excitations remain in the continuum above the finite threshold $1/\tau_F$.

\section{Conclusions}
\vspace{-0.4cm}

The main lesson of this work is that extended hydrodynamic theories \cite{Jou_Extended,Liu1986,Muller_book,Salazar2020,GavassinoFronntiers2021,GavassinoNonHydro2022,Amoretti2026Price}, such as Cattaneo \cite{cattaneo1958}, Israel--Stewart \cite{Israel_Stewart_1979}, and radiative moment closures \cite{Farris2008,Sadowski2013,GavassinoRadiazione}, can be considerably better justified, and considerably more accurate, than is often assumed \cite{Geroch1995,LindblomRelaxation1996,Geroch:2001xs,StephanovHydroPlus:2017ghc,Denicol_Relaxation_2011,Wagner:2023jgq}. Their validity does not necessarily require a parametrically large separation of timescales, a finite-dimensional slow sector, or a weakly-broken symmetry. What matters is the spectral structure of the collision operator on the Hilbert space relevant to the observables of interest. When the appropriate conditions are satisfied, a finite-dimensional quasihydrodynamic EFT exists, and its accuracy can be systematically improved by adding higher-order corrections without introducing additional degrees of freedom.

Models 1 and 2 illustrate why finite moment truncations can sometimes be justified from an EFT perspective, rather than regarded merely as phenomenological approximations. In Model 1, the relaxation gap is not parametric: the first omitted modes relax only three times faster than the retained current. Nevertheless, the quasihydrodynamic expansion converges within a finite domain, and already at second order it reproduces the exact kinetic dispersion relations with remarkable accuracy. 
More generally, any finite spectral gap separating a finite-dimensional slow sector from the rest of the spectrum is sufficient to guarantee a controlled EFT at sufficiently small frequencies and wavenumbers. Model 2 demonstrates a different possibility: the microscopic slow sector may be infinite dimensional, while only a finite number of its modes couple to the observables of interest. The remaining slow modes can then be excluded by restricting the Hilbert space on which the EFT is constructed. Importantly, neither model relies on an exotic collision mechanism specifically engineered to produce quasihydrodynamics. Small-angle scattering and scattering-dominated radiative transfer are standard kinetic processes, encountered in a variety of physical settings \cite{PomraningFokker-PLanck,PomraningFluxFokker,pomraning1973equations,mihalas_book,Caro1983,Levermore1984,Hu1993,Akcasu1998,Gomart2011,DemattVzquez2026}.

Model 3 provides a complementary example. Here, a finite-dimensional slow sector is parametrically separated from the rest of the spectrum, as in conventional quasihydrodynamics, but the origin of this separation is unconventional. By making momentum diffusion much less efficient at low energies than at high energies, we create a bottleneck that suppresses the exchange between left- and right-moving particles. Remarkably, the structure of the resulting collision operator guarantees that only one additional mode becomes parametrically slow: the particle current. This produces a Cattaneo EFT without appealing to the usual mechanisms involving weakly broken translation invariance or the relaxation of superfluid motion. Instead, quasiconservation emerges from the low dimensionality and the spectral properties of the kinetic operator (see \cite{AhnBaggioli:2025odk} for a similar phenomenon in a holographic model)

These results also suggest that the widespread use of the relaxation-time approximation as a representative model of kinetic theory \cite{RomatschkeCutsandPoles:2015gic,Kurkela:2017xis,HellerInsights:2020hnq,Bajec:2024jez,Brants:2024wrx,Bajec:2025dqm} may have obscured the range of quasihydrodynamic phenomena that kinetic equations can support. The simple spectral structure of RTA is not a generic property of kinetic theory. More structured collision operators can exhibit isolated slow eigenvalues, multiple relaxation scales, and dynamically decoupled sectors, allowing quasihydrodynamic EFTs to emerge in situations where the usual symmetry-based intuition offers little guidance. Rather than regarding extended hydrodynamics as an exceptional consequence of weakly broken conservation laws, it may be more fruitful to regard it as a natural (surprisingly frequent) consequence of the spectral complexity of real-world kinetic theory.

\section*{Acknowledgements}

This work is supported by a MERAC Foundation prize grant,  an Isaac Newton Trust Grant, and funding from the Cambridge Centre for Theoretical Cosmology.

\appendix

\bibliography{Biblio}

\label{lastpage}
\end{document}